\documentclass[letter,twocolumn]{jpsj3}
\usepackage{graphicx,amsmath,amssymb,bm}

\allowdisplaybreaks 

\usepackage{color}
\definecolor{Green}{rgb}{0,0.7,0}

\newcommand{\e}{ {\rm e}}

\newcommand{\ey}{ \vec{e}_y}
\newcommand{\ez}{ \vec{e}_z}
\newcommand{\tkz}{\tilde{k}_z}
\newcommand{\tkx}{\tilde{k}_x}
\newcommand{\bk}{ \bm{k}}

\newcommand{\bkD}{ \bm{k}_{0}}
\newcommand{\kDx}{ k_{0x}}
\newcommand{\kDy}{ k_{0y}}
\newcommand{\kDz}{ k_{0z}}

\newcommand{\ET}{ $\alpha$-(BEDT-TTF)$_2$I$_3$ \;}
\begin{document}
%-------------------------------
\title{
Berry Phase of Dirac Nodal Line Semimetal 
in Single-Component Molecular Conductor
}
\author{
Yoshikazu Suzumura\thanks{E-mail: suzumura@s.phys.nagoya-u.ac.jp}, 
 and  Ai Yamakage 
}
\inst{
Department of Physics, Nagoya University,  Chikusa-ku, Nagoya 464-8602, Japan 
}

\recdate {  \;\;\;\;\;\;\;\;\;\; }
%------
\abst{
 The Berry phase and curvature are studied for the Dirac nodal line 
  semimetal of single-component molecular conductor [Pd(dddt)$_2$].
Using two-band model on the basis of a tight-binding model, it is shown that the Berry curvature, $\bm{B}(\bk)$, 
exists along a loop of the nodal line, on which the  Berry phase is obtained 
from  surface integral of $\bm{B}(\bk)$. 
 The Berry phase, which is called  Zak phase,
  is also calculated from one-dimensional integral of the Berry connection 
    along a line  between two equivalent points of  boundaries of 
 Brillouin  zone. 
The possible experiment is discussed in terms of  the Berry phase.
  }

%\begin{document}

\maketitle

It has been known that Dirac electrons  appear  in condensed matter 
 when two neighboring bands such as the conduction and valence bands  degenerate  at a crystal momentum.\cite{Herring1937}
The Dirac fermion  which originally was found in the massive 
relativistic particle     is realized  in graphite
  as the massless Dirac electron, which 
  obeys the Weyl equation.\cite{McCure1956} 
 Since the discovery of such electron in graphene monolayer,
\cite{Novoselov2005_Nature438}
  several properties of two-dimensional 
Dirac electrons have been studied extensively.
The Dirac electrons, which exist as the bulk system and 
are located close to chemical potential,
 have been found in  organic conductor\cite{Katayama2006_JPSJ75} 
    \ET   and single-component molecular conductor\cite{Kato_JACS}
     [Pd(dddt)$_2$]
 where the former shows  two-dimensional Dirac electrons 
 and the latter shows  three-dimensional Dirac electrons 
 with Dirac  nodal line  semimetal.\cite{Kato2017_JPSJ}
Such nodal line has been studied due to unconventional loop 
 of the Dirac points.\cite{Murakami2007,Burkov2011,Yamakage2016_JPSJ,
Fang2016,Yang2018,Hirayama2018,Bernevig2018}
The noticeable phenomena  of Dirac electrons come from topological property  of   wavefunction,  which is known as  the Berry phase.\cite{Berry1984} 
Such topological property is fundamental 
for   the Hall conductivity,
\cite{Novoselov2005_Nature438}
 in addition to the energy dispersion of the Dirac cone, 
 which  takes a crucial  role for the almost temperature-independent conductivity.\cite{Suzumura2018_JPSJ_T}

The present paper focuses on the Dirac electrons in the molecular conductor, 
 which are  most promising candidate  to observe these properties 
 due to  chemical potential located close to the nodal line.
 Although the Berry phase\cite{Berry1984} for the Dirac point of \ET 
 was shown explicitly,\cite{Suzumura2011_JPSJ}  
 that of  [Pd(dddt)$_2$] is not yet clarified due to 
 non-coplanar three-dimensional loop.
Thus, we calculate  the Berry curvature in  [Pd(dddt)$_2$] using 
     a reduced Hamiltonian of two-band model, which is similar to 
       that of CaAgX.\cite{Yamakage2016_JPSJ} 

We consider a Dirac  electron system consisting of  $M$ molecular sites 
 per unit cell,   $M=8$ for [Pd(dddt)$_2$].\cite{Kato2017_JPSJ}
The Schr\"{o}dinger equation is given by 
%--------------    (1)  ----------------
\begin{align} 
 H(\bm{k}) |n(\bm{k})> = E_n |n(\bm{k})> \; ,
\end{align}
%--------------------------------------
 where  $\bm{k} =(k_x,k_y,k_z)$ is  a  three-dimensional  wave vector
 with a lattice constant taken as unity, and
 $n (= 1, 2, \cdots, M)$ denotes a band index. Quantities 
 $H(\bm{k})$,  $E_n(\bm{k})$ and  $|n(\bm{k})>$  are 
 the $M$ $\times$ $M$ matrix Hamiltonian with 
 the base of  8 molecular orbitals,   eigenvalue (band energy)  and  eigenfunction (wavefunction), respectively. 
The nodal line in the present system exists between 
 $E_4(\bk)$ and  $E_5(\bk)$,  which are the conduction and valence bands
  due to a half-filled band.  

 The Berry phase $\gamma_n$ is given by\cite{Berry1984} 
\begin{subequations}
% ----------------- (2a) ------------------------
\begin{eqnarray}
 \gamma_n = i \int_C \; d \; \bk \cdot <n(\bk)| \nabla_{\bk}|n(\bk)>
 = \int_S \bm{B}_n(\bk) \cdot d \bm{S}\; , 
\label{Berry_a}
 \end{eqnarray}
\begin{eqnarray}
% ----------------- (2b) ------------------------
\bm{B}_n(\bm{k})
   =  -  {\rm Im} \{ (\bm{\nabla}_{\bm{k}} < n(\bk) |)  \times 
  \bm{\nabla}_{\bm{k}}| n(\bk) > \} \;  , 
\label{eq:formula_1}
\end{eqnarray}
\end{subequations}
%--------------------------------
where 
 $<n(\bk)| \nabla_{\bk}|n(\bk)>$ and $\bm{B}_n(\bm{k})$ 
 denote the Berry connection and curvature, respectively.
 $C$ denotes a closed path and $S$ is the area enclosed by the path.

 We examine   $\bm{B}$ and $\gamma$, which denote $\bm{B}_4(\bk)$ and 
  $\gamma_4$, respectively, for the conduction band.
As shown later, the  Dirac nodal line is understood 
 by  $\gamma$, which is a scalar quantity given by 
 $\pm \pi$ or zero, but depends on $\bk$  related  to 
 the  path $C$.  
 Since   $\bm{B}(\bk)$  is essentially 
       determined  by $E_4$ and $E_5$,
 Eq.~(\ref{eq:formula_1}) is calculated by introducing 
 a 2 $\times$ 2  reduced Hamiltonian 
     $ H^{\rm red} (\bm{k})$ relevant to these two bands.  
The Hamiltonian $H$  is rewritten as 
  $H = H_0 + H_1$, where 
 $H_0$  consists of  two kinds of electrons  with different parity 
 and $H_1$ denotes  the interaction between them.
Reduced Hamiltonian, $ H^{\rm red} (\bm{k})$,
 has  matrix elements  ($i$=1,2 and $j$=1,2) given by  
%----------- (3) -----------------
 \begin{eqnarray}
&& H^{\rm red} (\bm{k})_{ij} =  h_{ij} 
 \nonumber \\
&& = <\bar{i}| ( H_0(\bk) + H_1(\bk) - H_0(\bm{G}/2)) |\bar{j}>
 \; ,
 \label{eq:matrix_element}
 \end{eqnarray}
%----------------------------
where  $|\bar{1}>$ and   $|\bar{2})>$  are eigenfunctions 
of the conduction and valence bands  
 at a TRIM (time reversal invariant momentum) $\bk = \bm{G}/2$ 
   and are   calculated from   
 $H_0(\bm{G}/2) |\bar{1}> =\bar{E}_1 |\bar{1}>$, and 
 $H_0(\bm{G}/2) |\bar{2}> =\bar{E}_2 |\bar{2}>$.
 Thus, the general form of the two-band model of the 
 2 $\times$ 2 reduced Hamiltonian,  $H^{\rm red} (\bm{k})$,
  is expressed  as  
%--------------    (4)  ----------------------
 \begin{eqnarray}
 H^{\rm red} (\bm{k})  &   = &
 \begin{pmatrix}
f_3 + f_0 & f_1-if_2 \\
f_1+if_2 & - f_3 + f_0
\end{pmatrix}
 \nonumber \\ 
 &=&  f_0 + f_1 \sigma_1 + f_2 \sigma_2 + f_3 \sigma_3 \; ,
\label{eq:Heff_start}
\end{eqnarray}
%--------------------------------------------
where 
 $f_0 = (h_{11}+h_{22})/2$, 
 $f_3 = (h_{11}-h_{22})/2$, and 
$f_1 - i f_2 = h_{12} = f_{21}^*$.
The quantity $\sigma_j$ denotes the Pauli matrix and 
 the coefficient $f_j$ (= $f_j(\bm{k})$) is given as the function of 
 $\bm{k}$. 
 Although the term $f_0$ gives  the tilting of  the Dirac cone,
  $f_0$ is irrelevant to $\bm{B}$ and is discarded hereafter. 
Substituting Eq.~(\ref{eq:Heff_start}) into Eq.~(\ref{eq:formula_1}),
we obtain the Berry curvature as\cite{Suzumura2011_JPSJ}
%--------------    (5)  ----------------------
\begin{eqnarray}
 && \bm{B}(\bk)   =- \frac{1}{2E^3} \left(
 f_3(\bm{\nabla}_{\bk}f_1 \times \bm{\nabla}_{\bk}f_2) \right. 
 \nonumber \\
&& +   f_1(\bm{\nabla}_{\bk}f_2 \times \bm{\nabla}_{\bk}f_3)
   \left.
 +   f_2(\bm{\nabla}_{\bk}f_3 \times \bm{\nabla}_{\bk}f_1)
\right) ,
\label{eq:Berry_curvature}
  \end{eqnarray}  
%--------------------------------------------
where $E=\sqrt{f_1^2+f_2^2+f_3^2}$.

Here we mention a relation between the direction of the nodal line and that of 
the Berry curvature for a choice of constant $f_1(\bk)$, which is  examined 
 in the present paper. 
Since the Dirac nodal line consisting of Dirac point $\bkD$ is obtained by the intersection of two kinds of planes, 
%-----------------  (6)  -------------------
\begin{eqnarray}
   f_2(\bk) = 0, \;\; {\rm  and} \;\; 
   f_3(\bk) = 0,
\label{eq:nodalline_cond}
\end{eqnarray}
 the direction of the line is perpendicular to 
 both $\bm{\nabla}_{\bk}f_2(\bk)$ and $\bm{\nabla}_{\bk}f_3(\bk)$ , i.e., 
  $\bm{\nabla}_{\bk}f_2(\bk) \times \bm{\nabla}_{\bk}f_3(\bk)$
 at $\bk = \bkD$.
From Eq.~(\ref{eq:Berry_curvature}), we obtain 
$\bm{B}(\bk) \propto   \bm{\nabla}_{\bk}f_2(\bk) 
          \times \bm{\nabla}_{\bk}f_3(\bk)$  
  with $\bk= \bkD$.
Thus, the direction of $\bm{B}(\bk)$ 
 is either  parallel or antiparallel to the tangent of the line 
 and is determined by  the model as shown below.

In order to understand the basic behavior of the nodal line, 
  we first show the Berry curvature for CaAgP,\cite{Yamakage2016_JPSJ} 
 which consists of $P_z$ and $S$  orbitals with  different parity. 
 The corresponding two-band model  is given by 
    $f_2 = k_z, f_3 = C_0 +C_1k_z^2 + C_2 (k_x^2+k_y^2)$,
     $(C_0 > 0, C_1, C_2 <0)$, and  $f_1 = - \Delta$, 
     where   small $\Delta (>0)$ is introduced  to obtain finite $\bm{B}(\bk)$.
The energy is given by $\pm E$  where 
%-------------------  (7) -----------------------
 \begin{eqnarray}
   E = \sqrt{\Delta^2 + (vk_z)^2 +(C_0 +C_1k_z^2 + C_2 (k_x^2+k_y^2))^2 }\; . 
 \end{eqnarray} 
 From $f_2(\bkD)=f_3(\bkD)=0$,
 the nodal line on the coplanar plane of $k_{z}$ = 0 is obtained as 
  $\bm{k}_0 =(k_0 \cos \theta, k_0 \sin \theta, 0)$ 
 with $k_0 = \sqrt{|C_0/C_2|}$, which is in the $k_x$-$k_y$ plane.   
Substituting this $f_j$'s into Eq.~(\ref{eq:Berry_curvature}), 
$\bm{B}(\bk)$ is calculated as 
%-------------------  (8) -----------------------
 \begin{eqnarray}
 \bm{B}(\bk) &=& \frac{\Delta}{2 E^2} [ 
    (0, 0, v)\times (-2|C_2|k_x, -2|C_2|k_y, 2 C_1k_z)]
                     \nonumber \\
  & & = \frac{v |C_2| \Delta}{E^{3/2}} (k_y, - k_x, 0) 
 = |\bm{B}(\bk)| \vec{e}_{\perp}\; , 
\label{eq:curvature_CaAgP}
\end{eqnarray}
where $\vec{e}_{\perp} = (\sin \theta, -\cos \theta, 0)$. 
It is found that $\bm{B}(\bk)$ is also in the $k_x$ - $k_y$ plane and 
 $\bm{k}_0 \cdot \bm{B}(\bk) = 0$. 
By introducing $\delta \bm{k} = \bm{k} - \bm{k}_0 =$ 
 $(\rho \cos \theta, \rho \sin \theta, k_z)$, 
the Berry phase $\gamma$, which is obtained from  
 two-dimensional integral perpendicular to $\vec{e}_{\perp}$,
  is calculated as   
%------------- (9) -----------------
 \begin{eqnarray}
 \gamma (\bkD)  =  \int_{- \infty}^{\infty} d \; k_z 
        \int_{- \infty}^{\infty} d \; \rho \; \; \bm{B} \cdot \vec{e}_{\perp}
\;
   = \pi  \; .
\label{eq:Berry_phase_CaAgP}
 \end{eqnarray}
Thus the Berry phase with $\pi$ 
 exists along the nodal line, 
in which the odd parity of $f_2(k_z)$ is  crucial.

%========  Fig 1 ================
\begin{figure}
  \centering
\includegraphics[width=8cm]{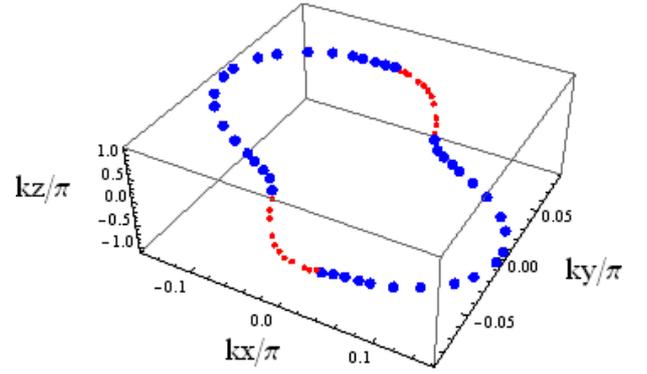}
    \caption{(Color online)
Nodal line of Dirac point at  $P$ = 8.0 GPa
in an extended  Brillouin zone, which 
 is obtained from 
 the original 8 $\times$ 8 Hamiltonian.\cite{Kato2017_JPSJ}
The chemical potential  ($\mu = 0.5561$ eV) exists at four Dirac points 
 $\bkD /\pi = \pm (-0.024, \pm 0.071, 0.65)$, which are located 
 between the large and  small symbols corresponding to  
   $\delta E > 0$ and  $\delta E < 0$  
      with $\delta E = E_4(\bkD)-\mu$. 
\cite{Suzumura2017_JPSJ_T=0}
The energy variation on the nodal is given by 
 $ -0.0021 \mathrm{\, eV} < \delta E < 0.0008 \mathrm{\, eV}$  and   $\delta E = 0$  
  on the chemical potential.
 }
\label{fig1}
\end{figure}
%----------------------------------------------

%========  Fig 2 ================
\begin{figure}
  \centering
\includegraphics[width=6cm]{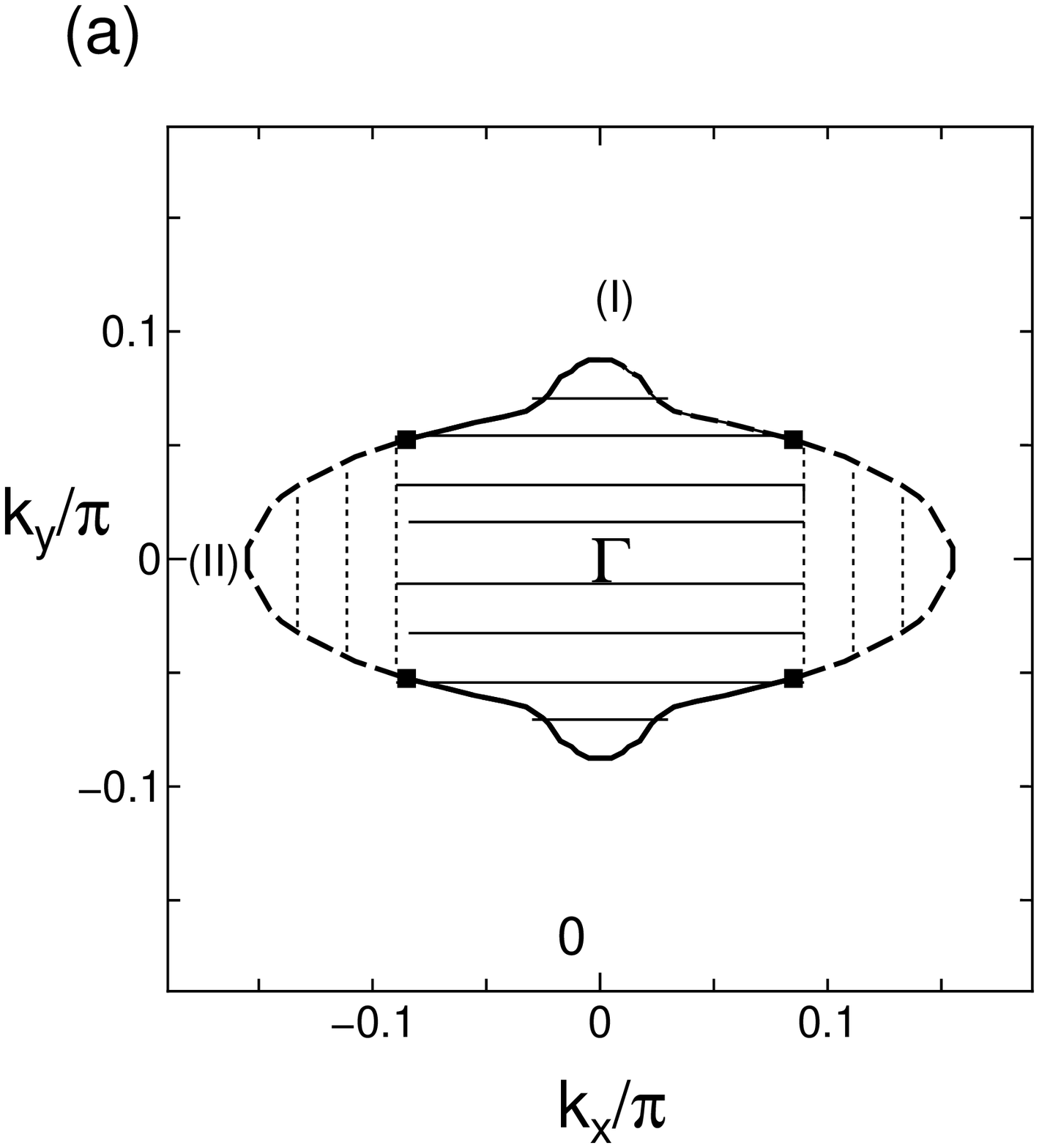}
\includegraphics[width=6cm]{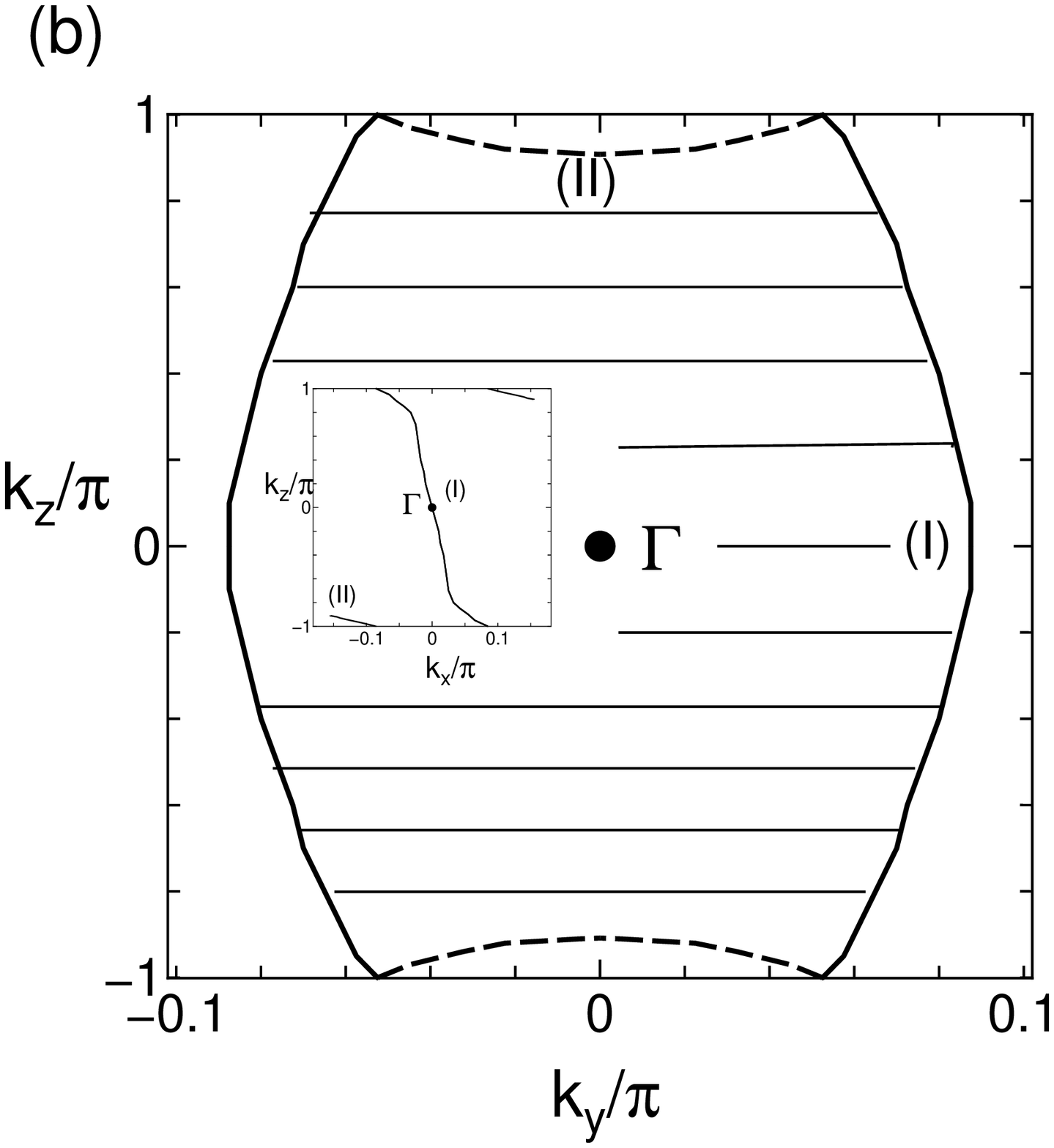}
    \caption{
Nodal line of Dirac point (solid and dashed lines) 
in the reduced  Brillouin zone corresponding to Fig.~\ref{fig1}, 
which is   projected 
on the $k_x$-$k_y$ plane (a) and 
on the $k_y$-$k_z$ plane (b). 
Typical Dirac points  are shown  
 for  $\bkD/\pi$ = (0,0.0875,0) (I) and (-0.15, 0, 0.908) (II). 
The closed square in (a) corresponds to Dirac point 
 at the zone boundary   
 given by  $\bk = \pm (-0.085, \pm 0.0525, 1)$. 
   The closed circle in (b) denotes $\Gamma$ point.  
 The shaded areas in (a) and (b) represent the Zak phase with $\pm \pi$, 
 calculated from  Eqs.~(\ref{Zak_Berry_z}) and
 (\ref{Zak_Berry_x}), respectively, where 
the dotted line comes from the  reduced zone. 
The inset in (b) denotes the nodal line on the $k_x$ - $k_z$  plane 
 where two lines degenerate due to symmetry with respect to $k_y=0$. 
 }
\label{fig2}
\end{figure}

 Now we examine the nodal line of  [Pd(dddt)$_2$]\cite{Kato_JACS} 
 consisting of  two layers and   8 molecular orbitals in the unit cell with  
  a base of 4 HOMO's ($|H1>, |H2>, |H3>, |H4>$) and 
  4 LUMO's ($|L1>, |L2>, |L3>, |L4>$)
 where HOMO (highest occupied molecular orbital)
  and LUMO (lowest unoccupied molecular orbital)
   have  different parities each other. 
  The Dirac point is calculated by 
  numerical diagonalization of  the 8 $\times$ 8 
 tight-binding Hamiltonian,\cite{Kato2017_JPSJ}
  where $H = H_0 + H_1$, $ H_0= H^{\rm HH} + H^{\rm LL}$,
  ($H^{\rm HH}$  separated from $H^{\rm LL}$) and 
 $H_1 = H^{\rm HL}$, (coupling between HOMO and LUMO).
The transfer energies between nearest-neighbor molecules are given by 
the interlayer ($z$ direction) and  intralayer ($x$ - $y$ plane)
 contributions. 
 The  matrix element is complex when 
  the same phase is taken for sites  in the unit cell.
Multiplying the base of both HOMO and LUMO by 
 a factor   
$(1, \e^{-i(k_x+k_y+k_z)/2)},\e^{-i(k_x+k_y)/2)}, \e^{-i(k_z)/2)})$
\cite{Kato2017_JPSJ}
, 
we obtain  $H_0$ being real and even function with respect to $\bk$  and that of $H_1$ being  pure imaginary and odd function with respect to $\bk$.
In Fig.~\ref{fig1}, the Dirac nodal line in an extended Brillouin zone 
 is shown in  three-dimensional $\bk$ space.
\cite{Kato2017_JPSJ,Suzumura2017_JPSJ_T=0}
The nodal line, which is elongated along the $k_z$ direction 
 and  symmetric with respect to $k_y = 0$,  touches the zone boundary at $k_z = \pm \pi$. 
It should be noted that the surface enclosed by the loop is not coplanar 
 since typical Dirac points   
  of   $\bkD/\pi = (0, \pm0.0875,0)$ (I), 
 $(\mp 0.015, 0, \pm 0.908)$ (II),  
 and  $ \pm (-0.085, \pm 0.053, 1)$ ( zone boundary)
 are not on the same plane. 
 The chemical potential 
 existing  at the Dirac point between the large  and  small symbols 
 suggests semimetallic state, 
 which gives rise to a large  response of the Dirac electron to
  the external field. 
Moreover large anisotropy of the Dirac cone,\cite{Kato2017_JPSJ} 
 which rotates along the nodal line 
 also gives  variety of transport measurements.

In order to examine analytically the Berry phase, 
 we use  Eqs.~(\ref{eq:matrix_element}) and (\ref{eq:Heff_start})
 where  
 $f_2 (\bk) = - f_2( - \bk)$, $f_3(\bk)= f_3(-\bk)$, 
  and $f_1(\bk)= - \Delta (\rightarrow - 0)$.  
 The Dirac points are symmetric with respect to 
 $\Gamma$ [= (0,0,0)] point and also 
 a plane of $k_y=0$. 
 The Dirac points of   Figs.~\ref{fig2}(a) and  \ref{fig2}(b), 
 which are  projected from Fig.~\ref{fig1},  
 show  difference  in magnitude  compared with  those 
  obtained from Eq.~(\ref{eq:nodalline_cond}), where 
 $f_{j}(\bk)$ is calculated by perturbational method.
The nodal line similar to Figs.~\ref{fig2}(a) and  \ref{fig2}(b) 
could be obtained by regarding 
  $f_{j}(\bk)$ as a renormalized one with keeping the parity, 
  which comes from   higher order in perturbation as was usually  treated.
\cite{Kato2017_JPSJ,Yamakage2016_JPSJ} 
Thus we  examine  the topological property of the Berry phase 
 using the  analytical treatment of such two-band model. 
Since the nodal line is complicated, we mainly examine two cases  
with Dirac points given by 
 $\bkD/\pi = (0, \pm 0.0875, 0)$ (I), and  
  $\bkD/\pi = \pm (- 0.015, 0, 0.908)$, (II)
 and the case being slightly away from (I), i.e., 
$\bkD/\pi \simeq (0, \pm 0.0875, 0)$ (III).

For the case (I), the reduced model on the basis of TRIM  at the 
$\Gamma$ point  is given by 
 $f_2 = - C_1 k_x$,  
$f_3 = C  - C_2 k_x^2 -C_3 k_y^2$, ($C, C_1, C_2, C_3 > 0$),
 $f_1 = - \Delta$, and $k_z=0$. 
In this case , $\bkD = (0, \kDy, 0 )$ with $\kDy = (C/C_2)^{1/2}$.
Expanding $k_y =\kDy + \delta k_y$,  
 $\bm{B}(\bk)$ close to $\bkD$ is calculated as  
%------------ (10) --------------------------
 \begin{eqnarray}
 \bm{B}_{\rm I}(\bk) &=& \frac{  C_1 C_3 \kDy \Delta}
   {(\Delta^2 + (C_1k_x)^2 + (2 C_3 \kDy \delta k_y)^2)^{3/2}}
       \ez  
       \; , %\nonumber \\
    \label{eq:curvature_I}
 \end{eqnarray}
 where $\ez = (0, 0, 1)$ and $\bm{B}_{\rm I}(\bk)$ is perpendicular to 
 the $k_x$ - $k_y$ plane.  
The Berry phase $\gamma$ obtained from  
 two-dimensional integral perpendicular to 
 $\ez$  
  is calculated as   
%-------- (11)----------------------
 \begin{eqnarray}
 \gamma (\bkD) =  \int_{- \infty}^{\infty} d \; k_x 
        \int_{- \infty}^{\infty} d \; k_y \; \; \bm{B}_{\rm I} \cdot  \ez \;
   =  \pi \kDy/ |\kDy| \;  \; ,
\label{eq:Berry_phase_I}
 \end{eqnarray}
which gives  $  \pm \pi$.

For the case (II), the reduced model around   the 
$Z$  $[ = (0,0,\pi)]$ point  is given by 
 $f_2 = - C_1 \tkz$,  
 $f_3 = C + C_2 (\tkz)^2 - C_3 k_x^2$, ($C, C_1, C_2, C_3 > 0$), 
and $f_1 = - \Delta$ where $k_y = 0$ and  
 $\tkz = k_z + {\rm sgn}(k_x)\pi - 0.16 k_x$. 
In this case, $\bkD = (\kDx, 0,\kDz)$ with $\kDx = \pm (C/C_3)^{1/2}$. 
Expanding $k_x = \kDx + \tkx$,  
 $\bm{B}_{\rm II}(\bk)$ close to $\bkD$ is calculated as 
%--------------- (12) -----------------------
 \begin{eqnarray}
 \bm{B}_{\rm II}(\bk) = \frac{ C_1 C_3 \kDx \Delta}
   {(\Delta^2 +(C_1 (\tkz-\kDz))^2 + (2C_3\tkx)^2)^{3/2}}
       \; \ey \; ,
    \label{eq:curvature_II}
 \end{eqnarray}
 where $\ey = (0, 1, 0)$, and $\bm{B}_{\rm II}(\bk)$ 
  is perpendicular to the $k_z$ - $k_x$ plane.  
The Berry phase $\gamma$, which is obtained from  
 two-dimensional integral perpendicular to 
 $\ey$, 
  is calculated as   
%----------- (13) -------------------
 \begin{eqnarray}
\gamma (\bkD) =  \int_{- \infty}^{\infty} d \;\tkx \; 
        \int_{- \infty}^{\infty} d \; \tkz \; \; \bm{B}_{\rm II}\cdot  \ey \;
   = \pi \kDx/ |\kDx| \;  ,
\label{eq:Berry_phase_2}
 \end{eqnarray}
 showing $ \pm \pi$.

Further, we examine the Berry curvature for the case (III), 
 which is  slightly away from  the case (I). 
Our perturbational calculation\cite{Kato2017_JPSJ} 
 gives  
$f_2 = - C_1 k_x -C_y k_y^2k_z$, 
$(C_1,  C_y >0)$,
$f_3= C -C_2k_x^2 -C_3k_y^2$, 
$(C, C_2, C_3 > 0)$, and 
$f_1 = - \Delta$
where 
 $\bkD =(\kDx,\kDy,k_z)$
with $\kDx \simeq - k_z(C_y C/C_1C_3)$ and  $ \kDy \simeq (C/C_3)^{1/2}$
for  small $k_z$. 
Such  Dirac point corresponds to  the accidental degeneracy since 
 the line given by $f_2(\bk)=0$ is irrelevant to the crystal symmetry.
The Berry curvature close to  
$\bk= \bkD$ is obtained as
$\bm{B}_{\rm III}(\bk) \propto 
 ( -C_3C_y (\kDy)^3,\;  C_2C_y \kDx (\kDy)^2,\;  
   C_1C_3 \kDy - 2 C_2 C_y \kDx \kDy k_z)$. 
 Note that  $\bm{B}_{\rm III}(\bk)$  at $\bkD= (0,\kDy,0)$ 
still has finite component along $k_x$ being consistent with the direction of 
the  nodal line at $\bkD= (0,\kDy,0)$. 
 However Eq.~(\ref{eq:Berry_phase_I}) still holds 
 since the total flux of  $\bm{B}_{\rm III}(\bk)$ 
on the $k_x$-$k_y$ plane is the same as 
 that of  $\bm{B}_{\rm I}(\bk)$.
These  results suggest  $\gamma = \pm \pi$ 
 for arbitrary $\bkD$ on the line, which is obtained 
 by substituting 
$ E = \sqrt{ \Delta ^2 + (\bm{\nabla}_{\bk}f_2 \cdot  \delta \bm{k})^2 
   + (\bm{\nabla}_{\bk}f_3 \cdot  \delta \bm{k})^2}$ 
 into Eqs.~(\ref{eq:Berry_curvature}) and (\ref{Berry_a}) 
  with $\delta \bm{k} = \bk-\bkD$ being  perpendicular to the nodal line. 
Thus, it is expected that the Berry phase rotates along the nodal line 
anti-clockwise around the $k_z$ axis in Fig.~\ref{fig1}.

Here we show the Berry phase obtained  from another point of view,
  which is  known as the Zak phase.
The Zak phase was originally obtained for one-dimensional periodic band.
\cite{Zak1988}
 In the present three-dimensional band, 
  such an idea is extended by choosing a line 
     connecting  two points of   $(k_x,k_y,-\pi)$ and  $(k_x,k_y,\pi)$, 
         which have the same wavefunction 
              due to $H(k_x,k_y,-\pi) = H(k_x,k_y,\pi)$.
Noting that the Dirac point or nodal line exists for the product of 
  the parity with all the TRIM's being $-1$,
\cite{Fu2007_PRB76} 
  the equivalence of  the Zak phase  and the Berry phase 
  is shown as follows.\cite{Kim2015}  
  When the parity of the $\Gamma$ point is different from 
  (the same as) 
  that of  the $Z$ point,  
   the  line integral of the Berry connection 
      between  $(0, 0,-\pi)$ and  $(0, 0,\pi)$  gives $\pm \pi$ (0) 
        due to  the inversion symmetry.\cite{Kim2015} 
Then, for  a closed path connecting 4 points 
 given by 
$(0, 0,-\pi)$ (A), $(0, 0,\pi)$ (B),
$(\pi, 0, \pi)$ (C), and $(\pi, 0, -\pi)$ (D),
 one finds  that the line integral gives 
  $\pm \pi$ for AB, 0 for BC+DA due to 
 the zone boundary, and 0 for CD due to both  C and  D having the same parity 
 as X $(\pi, 0,0)$. 
Therefore 
 the  Zak phase given  by AB is equal to the Berry phase given by  ABCD. 
Based on this argument, we obtain  the Zak phase with $\pm \pi$ 
 when  the line with  arbitrary $(k_x,k_y \pm \pi)$ exists 
   in the area including  the $\Gamma$ point. 
Thus, in terms of the Berry connection, 
 these phases  can be written explicitly as
%--------------- (14) -----------------------
\begin{subequations}
\begin{eqnarray}
 \gamma_z (k_x,k_y)   
&= & i \int_{-\pi}^{\pi} d \;  k_z\; <n(\bk)|\nabla_{k_z}|n(\bk)>
   \; , \nonumber    \\
   \label{Zak_Berry_z}
                            \\
 \gamma_x (k_y,k_z)   
&= & i \int_{-\pi}^{\pi} d \;  k_x\; <n(\bk)|\nabla_{k_x}|n(\bk)> 
       \; , \nonumber \\
   \label{Zak_Berry_x}
 \end{eqnarray}
\end{subequations}
which are  line integrals along a straight path connecting 
 two points of 
  $(k_x,k_y,\pm \pi)$ and   $(\pm \pi,k_y,k_z)$ 
 for $\gamma_z (k_x,k_y)$ and   $\gamma_x (k_y,k_z)$, respectively. 
 In Figs.~\ref{fig2}(a) and \ref{fig2}(b),
 the Zak phase   becomes   $\pm \pi$  
   in the region including the $\Gamma$ point (shaded area)
    and zero outside. 
 The dashed line comes from a nodal line of 
  the extended zone.
 As shown in the inset of Fig.~\ref{fig2}(b), there is no area 
  for $\gamma_y(k_z,k_x) = \pi$ due to the inversion symmetry of
 the band with respect to $k_y=0$.   
There are common features between these two Zak phases since 
the parity of the $\Gamma$  point is different from  the other TRIM's,
 e.g., Z and X points.\cite{Kato2017_JPSJ}

We comment on the recent paper by Liu et {\it al.},\cite{Liu2018}  
  who obtained  the nodal line of [Pd(dddt)$_2$]
   using  the same crystal data  as ref.~\citen{Kato_JACS}.
Their  two-band Hamiltonian, where  
  the  coefficients  $g_1(\bk)$ and $g_3(\bk)$   correspond to 
     our $f_2(\bk)$ and $f_3(\bk)$, respectively,   
  also gives the Berry curvature  by adding a small potential to 
  $g_2(\bk) \rightarrow \Delta$, which  
      is clear  from  Eq.~(\ref{eq:Berry_curvature}). 
 The  nodal line in ref.~\citen{Liu2018} is located 
 close to the $\Gamma$ point 
   while that of the tight-binding model\cite{Kato2017_JPSJ} 
     exists in  the extended  zone.
Such geometry of the nodal line is also obtained 
in the previous DFT calculation,\cite{Tsumuraya2018}
 where a pair of Dirac points appear at the fractional coordinates 
 of $\bkD/2\pi$ = (0.0, $\pm$ 0.085, 0.0) 
along the $\Gamma$-Y direction and the loop extends 
 to those of (-0.1960, 0.000, $\pm$ 0.3875) 
in the $k_x$ - $k_z$ plane. 
This shows a difference 
  between the DFT calculation and the tight-binding model 
based on the extended H\"uckel calculation.
Further the DFT calculation shows  a  clear evidence  
 of the non-coplanar  nodal line, e.g., 
  a  Dirac point   
$ \bkD/2\pi = (-0.1310, \pm 0.0570, 0.200)$ is located away from 
 the same plane.\cite{Tsumuraya2018}
This non-coplanar behavior, which shares a common feature with the tight-binding mode,l\cite{Kato2017_JPSJ} 
 indicates the  most complicated origin of the accidental degeneracy.\cite{Herring1937} 
 Note that Eq.~(\ref{Zak_Berry_z}) is equal to the  Wannier charge center 
 in ref.~\citen{Liu2018}, which is obtained by 
 the summation of all the occupied bands.

Finally we discuss the possible experiment on   the nodal line 
 through the Berry phase.
The Landau level ($\propto \sqrt{n B}$) ( $n$ being an integer) 
 relevant to the  Dirac cone of Fig.~\ref{fig1} is expected 
 for  all the directions due to the loop and large anisotropy.   
\cite{Suzumura2018_JPSJ_T}
There is also a suggestion for   the surface  charge polarization 
  in the nodal line semimetal,
\cite{Murakami2017} 
 which is proportional to 
the area of the Zak phase, $S_{\rm Zak}$.
 From the shaded region in Fig.~\ref{fig2}(a) and \ref{fig2}(b),
  the ratio of $S_{\rm Zak}$ to that of each Brillouin zone 
 in the present case
 is given by $\simeq$ 0.01 and 0.06, respectively.

In summary,  we examined the Berry phase associated with the Dirac nodal line 
 of [Pd(dddt)$_2$], where the almost temperature-independent resistivity 
 was observed. The present nodal line with the three-dimensional loop
  provides the Berry curvature and the Zak phase 
 essentially  given by $\gamma_x(k_y,k_z)$, 
 which is compatible with the direction 
   of  the conductivity showing the typical property of the  Dirac electrons.
\cite{Suzumura2018_JPSJ_T} 

%-----------------------
\acknowledgements
%----------------------
One of the authors (Y.S.) thanks R. Kato and T. Tsumuraya for useful 
 discussions on the Dirac nodal line of [Pd(dddt)$_2$].

%========================================

\end{document}